\documentclass[aps,pre,reprint]{revtex4-2}

\usepackage{amsmath}
\usepackage{amsfonts}
\usepackage{amssymb}
\usepackage{physics}
\usepackage{graphicx}
\usepackage{mathtools}

\newcommand{\sS}{_\textsc{s}}
\newcommand{\sB}{_\textsc{b}}
\newcommand{\sG}{_\textsc{g}}
\newcommand{\SG}{^\textsc{g}}
\newcommand{\sI}{_\textsc{i}}
\newcommand{\sE}{_\textsc{e}}
\newcommand{\sP}{_\textsc{p}}

\newcommand{\sSB}{_\textsc{sb}}

\newtheorem{theorem}{Theorem}
\newtheorem{postulate}[theorem]{Postulate}

\begin{document}
   
   \title{A qubit strongly interacting with a bosonic environment: Geometry of thermal states}
   \author{Patrick Lee Orman}
   \author{Ryoichi Kawai}
   \affiliation{Department of Physics, University of Alabama at Birmingham, Birmingham, AL 35294, USA} 

\begin{abstract}
A standard theory of thermodynamics states that a quantum system in contact with a thermal environment relaxes to the equilibrium state known as the Gibbs state wherein decoherence occurs in the system's energy eigenbasis. When the interaction between the system and environment is strong, a different equilibrium state can be reached that is not diagonal in the system energy eigenbasis. Zurek's theory of einselection predicts that the decoherence takes place in the so-called pointer basis under the strong coupling regime, which can be viewed as continuous measurement of the system by the environment. The thermal state under the strong coupling regime is thus expected to be diagonal in the pointer states rather than energy eigenstates.  We have postulated that the thermals state in the strong coupling limit is a Gibbs state projected onto the pointer basis and have demonstrated this with a simple model of single qubit strongly interacting with a bosonic environment.
\end{abstract}

\date{\today}

\maketitle

\section{Introduction}

A standard theory of statistical thermodynamics tells us that a system in thermal equilibrium should be in the Gibbs state $\rho\SG = e^{-\beta H\sS}/Z\sS$ where $H\sS$ and $Z\sS$ are the system Hamiltonian and the partition function, respectively.  This was justified in various different ways. For example, the maximum entropy principle with an energy constraint is used in popular textbooks.  However, \emph{how} the system, starting from an arbitrary state, approaches the Gibbs state is still debated.  An isolated system does not reach a steady state under unitary dynamics unless it is in an energy eigen state. Thus it does not reach the Gibbs state in an exact sense.  The eigenstate thermalization hypothesis (ETH) is introduced to link the thermal equilibrium and the Gibbs state for systems with large degrees of freedom.\cite{Deutsch2018} On the other hand, if the system is in contact with environments, its dynamics becomes stochastic and a unique steady state emerges in which detailed balance is satisfied. The steady state is shown to be the Gibbs state if the system-environment interaction is sufficiently weak compared to the system energy.

When the system is reduced to a microscopic size, the coupling energy may be as large as the system energy and thus the standard theory of statistical thermodynamics fails.  A popular resolution to the strongly coupled system assumes that the total system (including the system and the environment) is in the Gibbs state $\rho\sSB = e^{-\beta(H\sS + H\sB + H\sI)}/Z$ and then, the state of the system can be obtained by tracing out the environmental degrees of freedom, $\rho\sS = \tr\sB \rho\sSB$, which has been written in a Gibbs-like form $\rho\sS \equiv e^{-\beta H\sS^*}/Z\sS^*$ where an effective Hamiltonian is defined as  $H\sS^* \equiv -\frac{1}{\beta} \ln \left[\frac{\tr\sB\{e^{-\beta (H\sS+H\sB+H\sI)}\}}{\tr\sB\{e^{-\beta (H\sB+H\sI)}\}}\right]$.  This effective Hamiltonian, also known as the Hamiltonian of mean force, has been used to investigate non-equilibrium thermodynamics in th strong coupling regime.\cite{Gelin2009,Campisi2010,Hilt2011,Esposito2015,Seifert2016,Jarzynski2017,Miller2017,Strasberg2019}

One of the most notable features of the Gibbs state is that coherence between energy eigenbasis is completely lost and that  approaching to the Gibbs state necessarily involves decoherence in the energy eigestates.  Quantum master equations based on the Born-Markovian approximation, which is valid only when the coupling is weak, show that such decoherence indeed takes place, and the system reaches the Gibbs state.\cite{Breuer2002}  However, when the coupling is strong, the Hamiltonian of mean force and other approaches based on non-Markovian dynamics suggest that the thermal state is not necessarily diagonal in the energy eigenbasis and that decoherence may take place in a different basis.\cite{Mori2008,Genway2012,Lee2012,Cai2014,Xiong2015,Vega2017}  However, a general expression of such thermal states is not known other than the Gibbs-like state based on the Hamiltonian of mean force.

Decoherence of small quantum systems is intensively investigated for the development of quantum computers and also for the quantum-to-classical transition.\cite{Schlosshauer2007,Buchleitner2009}   It has been shown that in the strong coupling regime, coherence between so-called pointer states is lost due to quantum entanglement between the system and the environments.\cite{Zurek1981,*Zurek1982,*Zurek2003,Mensky1997,*Mensky1998,Mensky2000} Thus the density matrix of the thermal state is expected to be diagonal in the pointer basis rather than the energy eingenbasis.  However, the actual value of the diagonal elements are not known. Our goal is to find them based on the Zurek's theory of environment-induced superselection (einselection).

The decoherence process can be viewed as projective measurement done by the environments,\cite{Mensky2000} which led us to a postulate that the Gibbs state is projected onto the convex hul of the pointer basis due to the continuous measurement by the environments.  Numerical simulations for a pair of qubits interacting strongly with two separate heat bath supported the postulate.\cite{Goyal2019}
In the present paper, we show that the dynamics of a single qubit strongly interacting with a bosonic environment is consistent with the postulate without ambiguity.
This paper is organized as follows. We first present a summary of the postulates in next section.  In section \ref{sec:models}, a simple model and a numerical method are introduced.  Then, the results and discussions will follow.

\section{Postulates}\label{sec:postulates}

When Zurek tried to develop a theory of quantum measurement processes, he introduced a concept of \emph{einselection} in which a system interacting with environment loses coherence in a particular basis set selected by the environments. The theory is not limited to the quantum measurement processes but also applicable to thermalization processes involving decoherence.  Here we use the same approach to investigate thermalization of a quantum system strongly interacting with an environment.

Consider a quantum system S and an environment (thermal bath) B.  Their Hamiltonians are denoted as $H\sS$ and $H\sB$, respectively.  They interact through a coupling Hamiltonian $\lambda V\sSB$ where $\lambda$ indicates the strength of the coupling.  We assume that the system asymptotically approaches a unique steady state $\rho\sS^*$ as time $t$ goes to infinity.  Although it has been reported that certain types of environments allow multiple steady states, we exclude such special cases.  Furthermore, we assume that  the steady state is the Gibbs state when the coupling is sufficiently weak, as predicted by the Born-Markovian master equations.

The thermalization to the Gibbs state involves decoherence in the energy eigenbasis $\ket{e_i}$.  Starting from an arbitrary state $\rho\sS(0)$, thermalization takes the system toward a convex hull $\Sigma\sE = \left\{ \rho=\sum_i Q_i \dyad{e_i} ;\, Q_i \ge 0 \wedge \sum_i Q_i = 1 \right\}$ in the Liouville space. Every point inside $\Sigma\sE$ corresponds to a mixed state diagonal in basis $\ket{e_i}$ and pure states correspond to the extreme points of the hull.\cite{Bengtsson2107}  When the system reaches $\Sigma\sE$, coherence among the basis $\ket{e_i}$ is completely lost.  The diagonal elements change until detailed balance is achieved.   The final state is the Gibbs state. In this picture, the thermalization involves two distinct processes, decoherence and energy thermalization.

When the system-environment coupling is strong, energy thermalization is significantly altered since the coupling can store a large amount of energy.  The Hamiltonian of mean force attempts to find a new thermal state by constructing an effective Hamiltonian.  It has been overlooked that the decoherence process is also strongly affected by the strong coupling.  The main cause of the decoherence is now quantum entanglement between the system and the environment.  The standard theory of decoherence suggests that coherence takes place in a basis set determined by the coupling operator $V\sSB$, the process known as environment-induced superselection or \emph{einselection}.\cite{Zurek1981,*Zurek1982,*Zurek2003}  We shall call such a basis set the pointer basis $\ket{p_i}$.  It is also suggested this decoherence process can be viewed as quantum measurement by the environment.\cite{Mensky1997,*Mensky1998,Mensky2000} Our postulates are based on this interpretation of the decoherence.

\begin{figure}
   \includegraphics[width=3.0in]{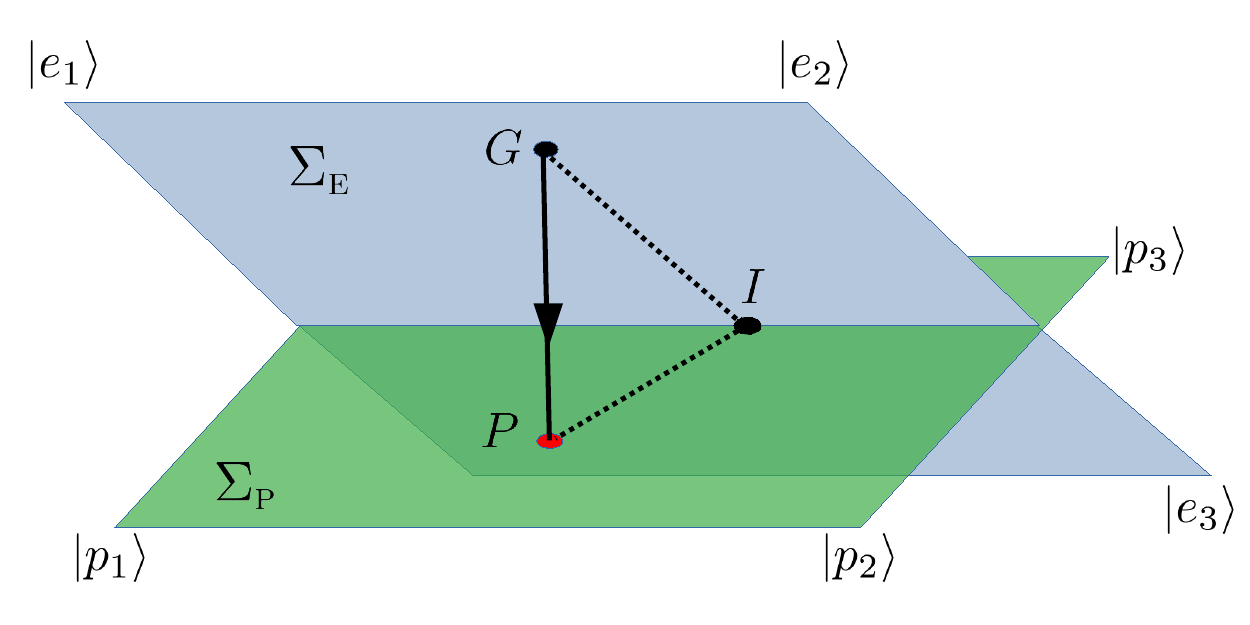}
   \caption{Schematic representation of Postulate \ref{thm:postulate1}. The Gibbs state on the convex hull $\Sigma\sE$ is projected onto another convex hull $\Sigma\sP$.  As the coupling strength increases, the steady state deviates from the Gibbs state ($G$) along the projection line toward the pointer limit ($P$). The maximal entropy state ($I$) is located on the intersection of the two convex hulls.  Noting that $P$ is closer to $I$ than $G$, the entropy increases as the steady state moves toward the pointer limit.}\label{fig:projection}
\end{figure}

Suppose that the system is in the Gibbs state ($G$ in FIG. \ref{fig:projection}). Zurek's einselection theory suggests that when the coupling is strong, the environment effectively measures a quantity of the system whose eigenvectors are $\ket{p_i}$ as suggested by Zurek. Assuming the measurement is projective, the Gibbs state $G$ on $\Sigma\sE$ is projected to $P$ on another convex hull $\Sigma_\textsc{p} = \left\{ \rho=\sum_i P_i \dyad{p_i};\, P_i \ge 0 \wedge \sum_i P_i = 1 \right\}$ as shown in FIG. \ref{fig:projection}.  We shall call $P$ the pointer limit. The projection line $\overline{GP}$ is ``perpendicular'' to $\Sigma_\textsc{p}$, meaning that the diagonal elements in the pointer basis are invariant along it. This consideration strongly suggests that the steady state shifts from the Gibbs state toward the pointer limit along the projection line as the coupling strength increases.  Based on this idea, we have proposed the following postulates.\cite{Goyal2019}

\nopagebreak
\begin{postulate}\label{thm:postulate1}
   At the strong coupling limit ($\lambda \gg 1$) the steady state density is given by, 
\begin{equation}
   \rho\sS(t)  \xrightarrow{t \rightarrow \infty} \rho\sP \equiv \sum_i  \dyad{p_i} \rho\sS^\textsc{g} \dyad{p_i}.
\end{equation}
\end{postulate}

\begin{postulate}\label{thm:postulate2}
   For any coupling strength ($\forall \lambda >0$), the diagonal elements of the steady state density in the pointer basis is given by 
\begin{equation}
  \expval{\rho\sS(t)}{p_i} \xrightarrow{t \rightarrow \infty}  \expval{\rho\sS\SG}{p_i}.
\end{equation}
\end{postulate}

At present we do not have a rigorous proof of the postulates.  In the following section, we will show that the postulates appear to be valid for a qubit coupled to a bosonic environment.

\section{Model}\label{sec:models}

We consider a qubit with Hamiltonian
\begin{equation}\label{eq:Hs}
   H\sS = \frac{\omega_0}{2} \sigma_z\, ,
\end{equation}
where $\omega_0$ is the excitation energy of the qubit.  In the present numerical calculation, we assume $\omega_0=1$ and thus all energy is normalized by $\omega_0$.

The qubit is coupled to an infinitely large bosonic environment
\begin{equation}
   H\sB = \sum_j \omega_j a^\dagger_j a_j\, ,
\end{equation}
where $a^\dagger_j$ and $a_j$ are usual creation and annihilation operators, respectively for $j$-th mode $\omega_j$. 

The system and environment are coupled through a bi-linear form of Hamiltonian
\begin{equation}
   V\sSB = X\sS \otimes Y\sB.
\end{equation}
An arbitrary system operator $X\sS$ linearly interacts with the displacement of each boson mode through coupling constant $k_j$ as
\begin{equation}\label{eq:YB}
   Y\sB = \sum_j \nu_j (a^\dagger_j + a_j).
\end{equation}
We further assume that the spectral density of the environment is of the Drude-Lorenz type
\begin{equation}\label{eq:Drude}
   J(\omega) = \frac{2 \lambda \gamma \omega}{\omega^2+\gamma^2}\, ,
\end{equation}
where $\gamma$  and $\lambda$ are relaxation rate and overall coupling strength, respectively. 

The whole system is completely isolated, and its time evolution is determined by Liouville-von Neumann equation $\dv{t} \rho\sSB = -i \comm{H\sSB}{\rho\sSB}$, where $H\sSB = H\sS \otimes I\sB + I\sS \otimes H\sB + V\sSB$ is the total Hamlitonian. We will omit the identity operators $I\sS$ and $I\sB$ in the following expressions.

Taking the partial trace of the whole system over the environment Hilbert space, we find the state of system $\rho\sS = \tr\sB \rho\sSB$ satisfies the non-unitary time-evolution
\begin{equation}\label{eq:eom}
   \dv{t} \rho\sS = -i \comm{H\sS}{\rho\sS} -i \comm{X\sS}{\eta_1} 
\end{equation}
where $\eta_n = \tr\sB (Y\sB^n \rho\sSB)$.  The thermal state $\rho^*\sS$ is defined as a steady state and thus $\comm{H\sS}{\rho^*\sS}+\comm{X\sS}{\eta_1^*}=0$.

\section{The Gibbs state and Pointer Limits}

\begin{figure}
   \includegraphics[width=2in]{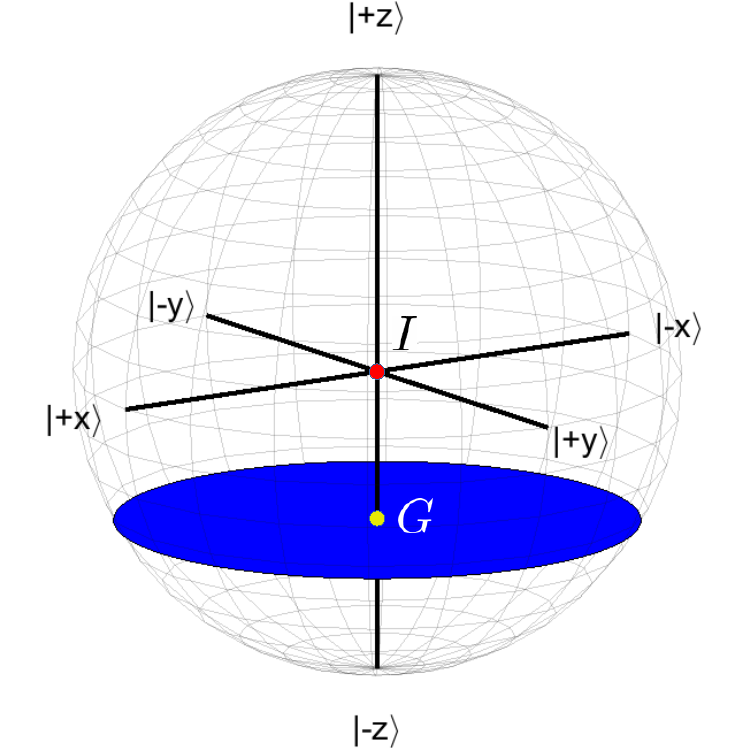}
   \caption{Gibbs state by the maximum entropy principle.  The constant energy plane (blue disk) is perpendicular to the $z$ axis.  The state with the highest entropy is located at the center of the sphere (I).  The maximum entropy state on the constant energy plane is the state nearest to the center, and thus G is the Gibbs state. It is necessarily on the $z$ axis or the convex hull $\Sigma_E$.}\label{fig:gibbs}
\end{figure}

Qubit states can be conveniently visualized with the Bloch sphere shown in FIG. \ref{fig:gibbs}.  Any state of a qubit is mapped to a point in the Bloch sphere. The point is specified by a radial vector $\va{r}$ (known as the Bloch vector), and the corresponding density matrix is expressed as $\rho\sS = \frac{1}{2} \left (I + \va{r}\cdot \va{\sigma} \right)$ where $\va{\sigma}$ is the vector Pauli operator.
With the system Hamiltonian \eqref{eq:Hs}, the Gibbs state can be written as
\begin{equation}\label{eq:bloch-gibbs}
\rho\sS\SG \equiv \frac{e^{-\beta H\sS}}{Z}=\frac{1}{2}\left[I\sS - \tanh\left(\frac{\beta \omega_0}{2}\right) \sigma_z \right]
\end{equation}
and the corresponding Bloch vector is given by 
\begin{equation}
\va{r}\sG = -\tanh\left(\frac{\beta \omega_0}{2}\right) \mathbf{e}_z.
\end{equation}
where $\mathbf{e}_z$ is a unit vector in the $z$ direction.  Noting that the $z$ axis is the convex hull $\Sigma\sE$, the Gibbs state is precisely on this convex hull and has completely lost coherency in the energy eigenbasis. 

The entropy of the system is measured by the von Neumann entropy $S = - \tr\sS \rho\sS \ln \rho\sS$, which is simply a function of radius $r\equiv \abs{\va{r}}$, 
\begin{equation}
S = \ln 2 -\frac{1}{2} \left[(1+r)\ln(1+r) + (1-r) \ln(1-r)\right].
\end{equation} 
which takes the highest possible entropy $\ln 2$ when $r=0$. The entropy decreases isotropically as $r$ increases.  The principle of maximum entropy states that the thermal equilibrium is the state with the highest entropy for a given energy.  In the current model, the constant energy surface is a plane perpendicular to the $z$ axis (indicated as the blue disk in FIG. \ref{fig:gibbs}).  The intersect of the $z$ axis and the plane is clearly the highest entropy point on the plane ($G$ in  FIG. \ref{fig:gibbs}).  Hence, the Gibbs state must be on $\Sigma\sE$.

From the dynamical point of view, the system, starting from any point in the sphere, is expected to thermalize to the Gibbs state.  Since  the $z$ axis is the convex hull $\Sigma\sE$ in the present model, any trajectory of $\rho\sS(t)$ moves toward the $z$ axis as the decoherence in the energy eigenbasis takes place.  The vertical drift of the trajectory along the $z$ axis  is due to energy relaxation through heat exchange with the environment. It turns out that this kind of trajectories involving decoherence toward the $z$ axis and energy thermalization along the $z$ axis,  is possible only when the coupling between the system and environment is very weak. As shown in the following sections, trajectories are quite different and the thermal state is not necessarily on the $z$ axis in the strong coupling regime. 

Based on Postulates \ref{thm:postulate1} and  \ref{thm:postulate2}, we can explicitly express the thermal state in the strong coupling limit.
As defined in the previous section, the pointer states $\ket{p_i}$ are the eigenvectors of the coupling operator $X\sS$. From Eq. \eqref{eq:bloch-gibbs} the diagonal elements of the pointer limit $\rho\sP$ predicted by Postulate \ref{thm:postulate1} are
\begin{equation}
 \mel{p_i}{\rho\sP}{p_i} = \frac{1}{2}\left[1 -\tanh\left(\frac{\beta\omega_0}{2}\right) \mel{p_i}{\sigma_z}{p_i} \right]\, ,
\end{equation} 
and off-diagonal elements all vanish.
If $\ket{p_i}$ are orthogonal to the eigenkets of $\sigma_z$, the pointer limit is simply $\rho\sP = \frac{1}{2} I\sS$ which carries the the maximum entropy.  From the geometrical point of view, the two convex hulls $\Sigma\sE$ and $\Sigma\sP$ are orthogonal, and the projection of the Gibbs state must be at their intersect. In the following section,  numerical simulation confirms these predictions.

\section{Numerical Experiments}

We solve Eq. \eqref{eq:eom} assuming that the whole system is initially in a product state $\rho\sS(0) \otimes \rho\sB\SG$ where the environment is in the Gibbs state $\rho\sB\SG$ at temperature $T$. The initial system state $\rho_s(0)$ can be any pure or mixed state.  Since we are interested in the steady state, starting with a product state does not cause a problem as long as there is one unique steady state.  Under these assumptions, Eq. \eqref{eq:eom} can be numerically solved using the method of hierarchical equations of motion (HEOM), which invokes neither Born or Markovian approximation.  HEOM is theoretically exact. However, in actual numerical implementation, some approximations such as truncation of infinite series are introduced, but the numerical errors are negligibly small.  (See Appendix \ref{apx:HEOM}.)

The numerical experiment was carried out for a variety of cases.  We tried various choices of coupling operator in the form of $X\sS = a_x \sigma_x + a_y \sigma_y + a_z \sigma_z$ with $\sum a_i = 1$. Here we show only two cases, $X\sS=\sigma_x$ and $X\sS=\left(\sigma_x + \sigma_z\right)/2$. For each choice of the coupling operator, we considered many different initial states including both pure and mixed states. All initial states reached the same steady state, suggesting that there is only one unique steady state for each $X\sS$.  The temperature is fixed at $T=1.5$ and the coupling strength $\lambda$ is varied from 0.01 to 5.

\subsection{Case I: $X\sS=\sigma_x$}

\begin{figure}
	\centering
	\includegraphics[width=3in]{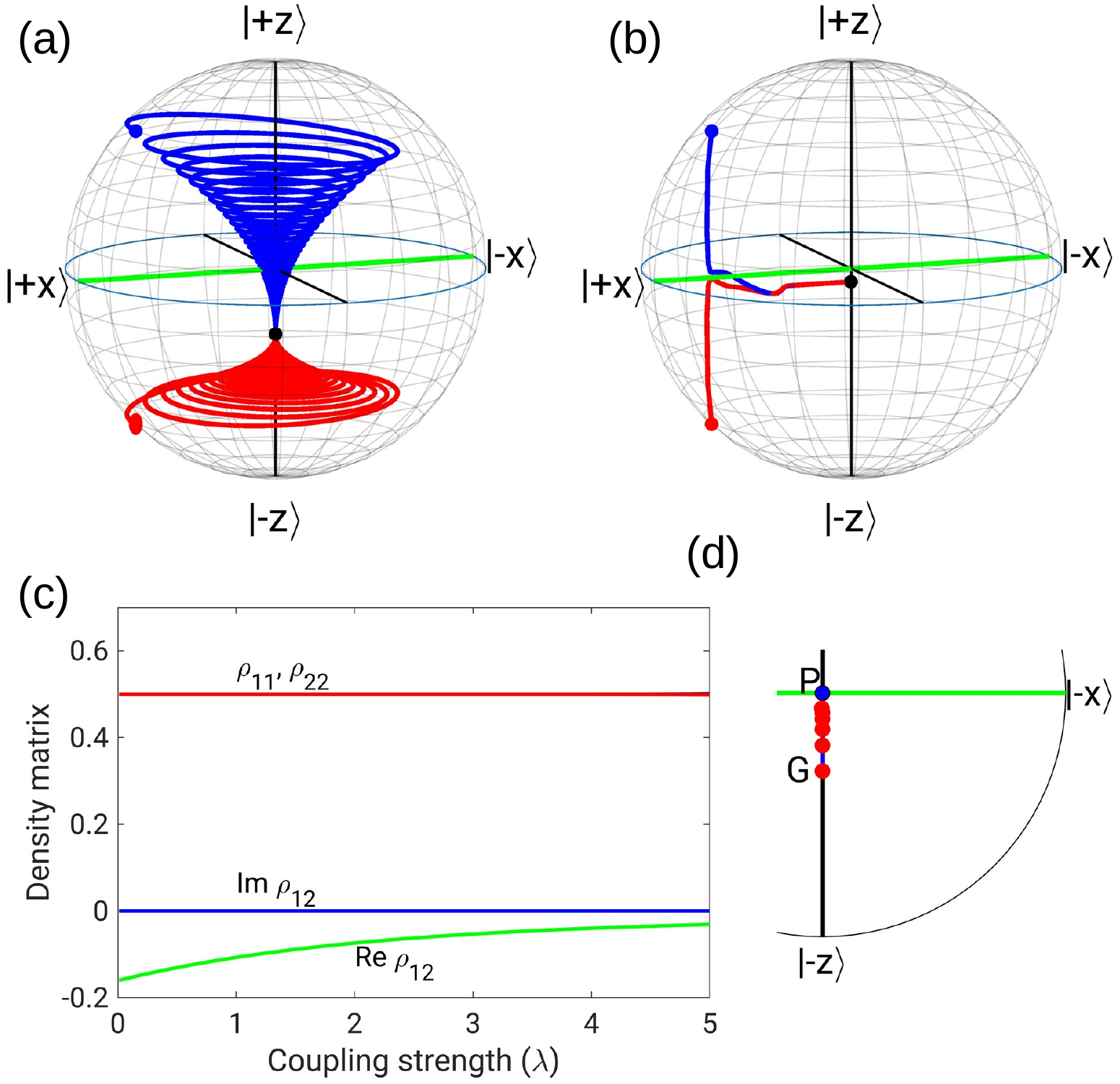}
	\caption{Steady state for $\sigma_x$ coupling.  (\textbf{a}) Two trajectories starting from different initial states are shown for weak coupling $\lambda=0.01$.  Both converge to the Gibbs state. (\textbf{b}) Two trajectories starting from the same initial state as \textbf{a} but with strong coupling $\lambda=5.0$.  The final steady state deviates from the Gibbs state and it is much closer to the highest entropy point. (\textbf{c}) The matrix elements of the system density in the pointer basis as function of the coupling strength $\lambda$.   The diagonal elements remain constant in consistent with postulate 2. (\textbf{d}) the transition of the steady state from the Gibb state to the pointer limit. The red circles show steady states for the coupling strength $\lambda=0.01, 1.0, 2.0, 3.0, 4.0, \text{ and } 5.0$ from G to P.  They follow the projection line from the Gibbs state (G) to the pointer limit (P).}\label{fig:sx-all}
\end{figure}

First, we consider a simple form of coupling, $X\sS=\sigma_x$.  The corresponding pointer states are $\ket{p_1} = \ket{x_+}$ and $\ket{p_2}=\ket{x_-}$ where $\ket{x_\pm}$ are the eigenkets of $\sigma_x$.  While the energy convex hull $\Sigma\sE$ is the $z$-axis,  the pointer convex hull $\Sigma\sP$ is the $x$ axis in the Bloch sphere.  As discussed in the previous section, $\Sigma\sE \perp \Sigma\sP$, and the pointer limit is exactly at the center of the Bloch sphere.  Here, Postulate $\ref{thm:postulate1}$ predicts that the thermal state is $\rho\sS^* = \frac{1}{2} I\sS$ which is independent of the temperature of the environment.  This thermal state carries the maximum possible entropy $S\sS=\ln 2$ and thus all information is lost to the environment.  In addition, Postulate \ref{thm:postulate2} predicts that $\expval{\rho\sS^*}{p_1}=\expval{\rho\sS^*}{p_2}=\frac{1}{2}$ for any value of the coupling strength, indicating that the steady state shifts from the Gibbs state toward the center of  the Bloch sphere along the $z$ axis.

In Fig. \ref{fig:sx-all}a and \ref{fig:sx-all}b, the trajectories of thermalization starting from two pure states, $\ket{\psi_1} = \frac{1}{\sqrt{2+\sqrt{2}}}\left(\ket{x_+}+\ket{z_+}\right)$ and $\ket{\psi_2} = \frac{1}{\sqrt{2-\sqrt{2}}}\left(\ket{x_+}+\ket{z_-}\right)$, are plotted.  Although we show only two trajectories, we tried many other initial conditions and all converged to the same steady state.  When the coupling is weak ($\lambda=0.01$), both trajectories spiral to the Gibbs state (FIG. \ref{fig:sx-all}a).  Decoherence (approaching the $z$ axis) and energy thermalization (drift along the $z$ axis toward the Gibbs state) happen simultaneously. There is no evidence that the pointer states play any role.  Starting with the same initial states, the trajectories under strong coupling ($\lambda = 5$) show rapid decoherence toward the $x$ axis, followed by slow drift to the thermal state on the $z$ axis.  The final steady state is much closer to the center than the Gibbs state. Figure  \ref{fig:sx-all}d illustrates that the thermal state deviates from the Gibbs state along the $z$ axis (projection line) toward the center (the pointer limit) as $\lambda$ increases in good agreement with Postulate \ref{thm:postulate2}.  

Since the final steady state is on $\Sigma\sE$($z$ axis), the decoherence looks like taking place in $\ket{z_\pm}$ basis.    However, in the current model, the pointer limit just happened to be at the center of the Bloch sphere where the density operator is diagonal in any basis set. We do not consider this slow drift to the $z$ axis as decoherence.   The initial rapid move toward $\Sigma\sP$ ($x$ axis) strongly indicates that the decoherence in $\ket{x_\pm}$ is induced by the system-environment entanglement.  
 
The density matrix in the pointer basis are plotted over coupling strength in Fig. \ref{fig:sx-all}c. As the Postulates predicted, the diagonal elements are both $\frac{1}{2}$ and remain constant as the coupling strength is varied.  The off-diagonal elements are clearly vanishing toward the strong coupling limit, indicating decoherence in the pointer basis.
The results exactly match the predictions made by the postulates.

\subsection{Case II: $X\sS=(\sigma_x+\sigma_z)/2$}

\begin{figure}
	\centering
	\includegraphics[width=3in]{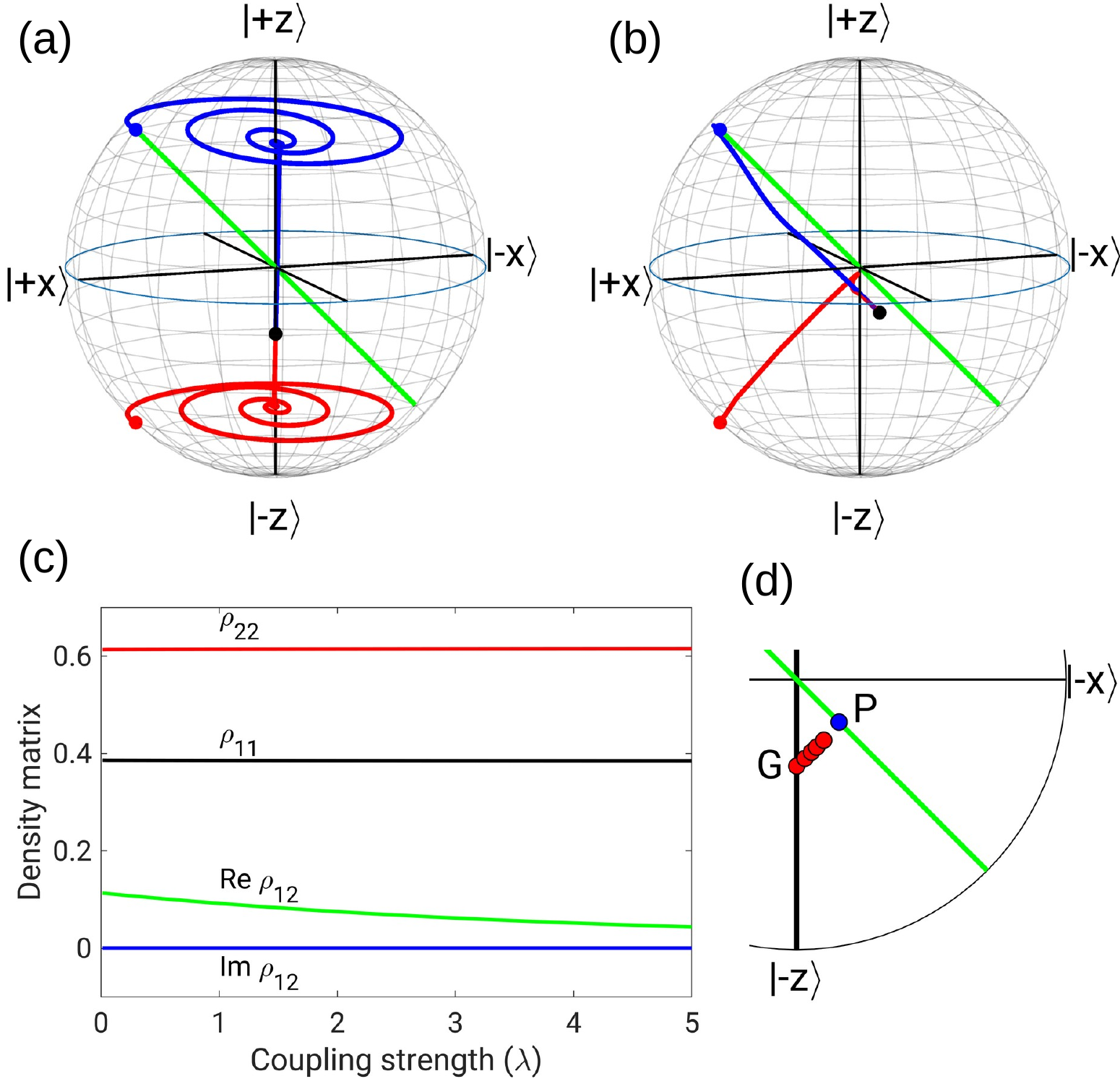}
	\caption{Steady state for $\sigma_x$ coupling.  (\textbf{a}) Two trajectories starting from different initial states are shown for weak coupling $\lambda=0.01$.  Both converge to the Gibbs state. The green line show the convex hull $\Sigma_p$ which seems taking no role.  (\textbf{b}) Two trajectories starting from the same initial state as \textbf{a} but with strong coupling $\lambda=5.0$. The final steady state deviates from the Gibbs state and it is no longer on the $z$ axis.  It has shifted toward $\Sigma_p$ (green line). (\textbf{c}) The matrix elements of the system density in the pointer basis as function of the coupling strength $\lambda$.   The diagonal elements remain constant in consistent with postulate 2. (\textbf{d}) the transition of the steady state from the Gibb state to the pointer limit along the green projection line. The red circles show steady states for the coupling strength $\lambda=0.01, 1.0, 2.0, 3.0, 4.0, \text{ and } 5.0$ from G to P.  They follow the projection line from the Gibbs state (G) to the pointer limit (P).}\label{fig:sxsz-all}
\end{figure}

Next we consider $X\sS =(\sigma_x+\sigma_z)/2$  where $\Sigma\sP$ is not perpendicular to $\Sigma\sE$.  The pointer states written in a mixed basis are 
\begin{equation}
\ket{p_1} = \frac{\ket{x_+}+\ket{z_+}}{\sqrt{2+\sqrt{2}}}, \quad
\ket{p_2} = \frac{\ket{x_+}-\ket{z_+}}{\sqrt{2-\sqrt{2}}}\, ,
\end{equation}
and Postulate \ref{thm:postulate1} predicts that the steady state density in the strong coupling regime is diagonal in this basis set.
The convex hull $\Sigma\sP$ is a line inclined from $\Sigma\sE$ by 45$^\circ$, and thus the pointer limit is no longer on the $z$ axis as shown in FIG. \ref{fig:sxsz-all}c.  The Bloch vector of the pointer limit (P in the figure) is given by $\va{r}\sP = -\frac{1}{2} \tanh(\beta/2) (e_x + e_z)$.  Based on Postulate \ref{thm:postulate2}, the steady state should lie on the projection line between G and P in Fig. \ref{fig:sxsz-all}c.

Two trajectories, one for weak ($\lambda=0.01$) and another for strong ($\lambda=5$) coupling, are plotted in Figs. \ref{fig:sxsz-all}a and \ref{fig:sxsz-all}b.  The weakly coupled system thermalizes to the Gibbs state as expected, but in a different way from the previous case. Rapid decoherence toward the $z$ axis happens before energy slow thermalization takes place.  For the strong coupling case, the initial state of the upper trajectory (blue trajectory in Fig. \ref{fig:sxsz-all}b)  happened to be on $\Sigma\sP$ and thus only slow gradual thermalization along $\Sigma\sP$ leads to the final point on the projection line.  On the other hand, the lower trajectory (red trajectory in Fig. \ref{fig:sxsz-all}b) show rapid decoherence to $\Sigma\sP$ and slowly converges to the point on the projection line.  As shown in  Fig. \ref{fig:sxsz-all}d, the steady state of various coupling strengths are all on the projection line and are moving toward the pointer limit as the Postulates claim.  Furthermore, the diagonal elements of the system density matrix in the pointer basis are completely independent of the coupling strength as plotted in Fig. \ref{fig:sxsz-all}c, in good agreement with Postulate \ref{thm:postulate2}.

\section{Discussions}

\begin{figure}
	\includegraphics[width=3in]{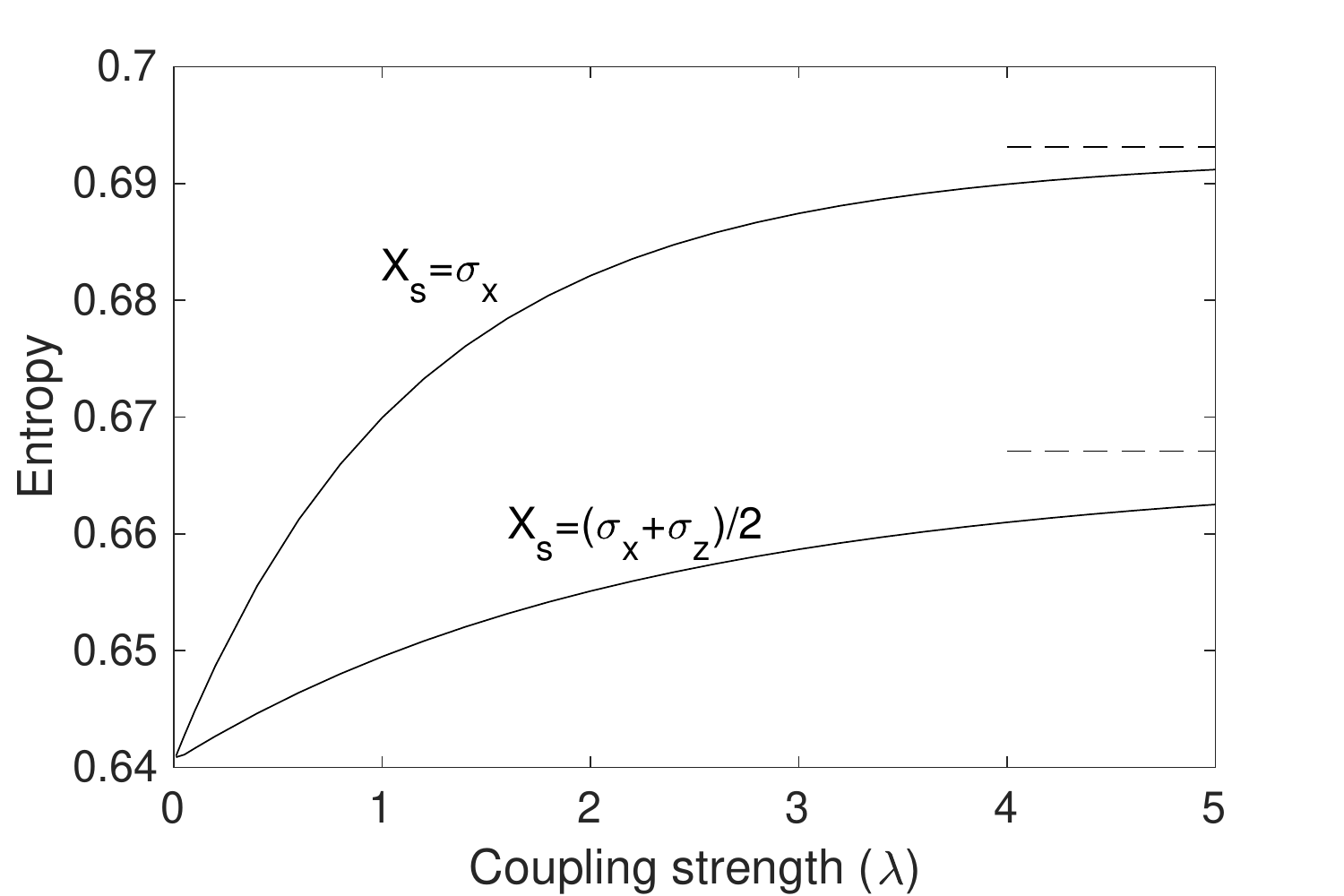}
	\caption{Increase in the qubit entropy toward the pointer limits (dashed lines) for both cases.}\label{fig:entropy}
\end{figure}

\begin{figure}
	\includegraphics[width=3in]{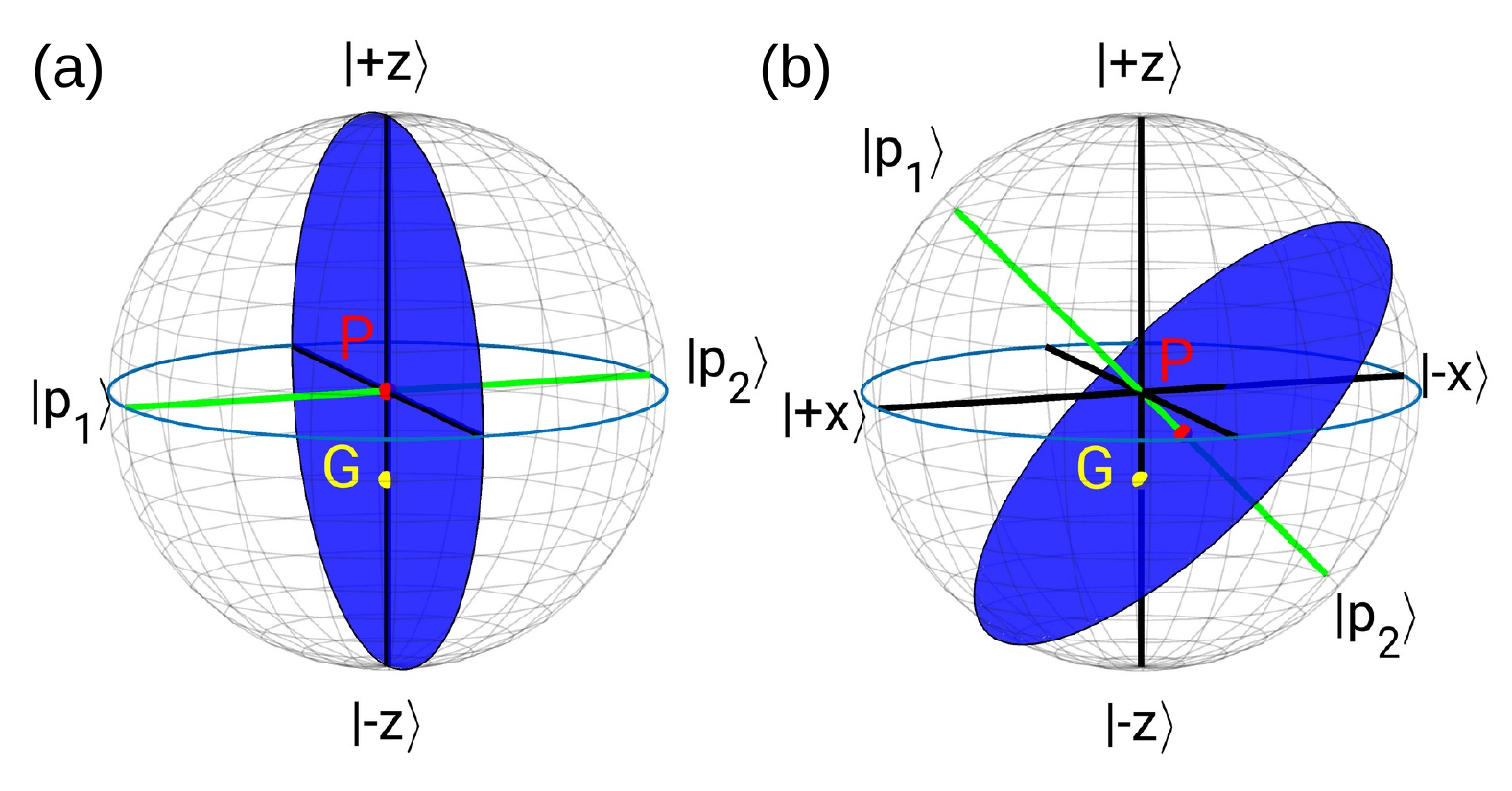}
	\caption{Maximization of entropy on the decoherence planes (blue disks) with $X\sS=\sigma_x$ (left sphere) and $X\sS=\frac{\sigma_x + \sigma_z}{2}$ (right sphere).}\label{fig:maxS}
\end{figure}

We have demonstrated the validity of Postulates \ref{thm:postulate1} and \ref{thm:postulate2} using a single qubit strongly interacting with a bosonic environment. The results of numerical simulations are all consistent with the Postulates.  Now we try to look at the postulates in context of entropy maximization.  Decoherence without change in the diagonal elements of the density matrix (dephasing) necessarily increases the entropy.  In fact, Fig. \ref{fig:entropy} shows that the entropy of the qubit increases toward the pointer limit as the coupling strength increases.
Then, Postulate \ref{thm:postulate1} is consistent  with the maximization of the system entropy on the decoherence plane involving the Gibbs state as shown in Fig. \ref{fig:maxS}.

The increase of entropy suggests that the ``effective temperature`` of the qubit is higher than the temperature of the environment.  In particular for $X\sS=\sigma_x$, the thermal state at the pointer limit  has infinite temperature.  It has been reported that heat flow through a pair of qubits between two heat baths vanishes under the strong coupling regime.\cite{Kato2015,Goyal2019}  It was said to be due to the quantum Zeno effect. However, heat vanishes only on certain cases, in particular when the coupling operator is $X\sS=\sigma_x$.  In this case, the effective temperature is much higher than that of heat bath, and heat flow from the thermal bath to the qubit becomes impossible.  On the other hand, the coupling $X\sS=(\sigma_x+\sigma_z)/2$ does not shut off the heat completely even in the strong coupling limit.  Hence, the loss of the heat is not simply due to the quantum Zeno effect.  In other words, the continuous measurement by the environment is complete with the former coupling but not with the latter.

In conclusion, decoherence in the pointer basis due to quantum entanglement with the environment  significantly affects thermodynamical processes such as quantum heat engines under the strong coupling regime.  The two Postulates appear to provide a useful physical picture of thermal equilibrium in the strong coupling limit.     The next step will be to construct a general theory of quantum thermodynamics in the strong coupling limit based on the pointer limit. 


\begin{acknowledgments}
   We would like to thank Janet Anders, James Cresser, Ala-Nissila, Sahar Alipour, Ali Rezakhani, and Erik Aurell for helpful discussion.  RK is grateful for NORDITA and Aalto University for their hospitality during his stay.
\end{acknowledgments}

\appendix
  
\section{HEOM}\label{apx:HEOM}
Tanimura and Kubo\cite{Tanimura1990} showed that the Feynman-Vernon influence functional can be obtained from an infinite set of ordinary differential equations called hierarchical equations of motion (HEOM).  It does not invoke any approximation, and its accuracy is solely determined by its numerical implementation.  Therefore, the results are in principle exact and non-Markovian effects, essential to the strongly coupled regime, are fully taken into account.   Since HEOM has been already used to study open quantum systems by other groups, we only briefly describe the method.\cite{Kato2015} 

When the environments are ideal Bose gases, the influence functional can be written in a super-operator form $ \overleftarrow{\mathcal{T}}e^{i \Phi(t,t_0)}$ where the super-operator is given  in the interaction picture by \cite{Breuer2002,Aurell2020}
\begin{equation}
   \begin{split}
   i \Phi(t,t_0) & =  \int_{t_0}^t \dd{s} \int_{t_0}^s \dd{s'} \mathcal{S}^{-}(s) \\
   & \quad \times \left \{i \kappa_i(s-s') \mathcal{S}^{-}(s') - \kappa_r(s-s') \mathcal{S}^{+}(s') \right\}
   \end{split}
\end{equation}
with super-operators $S^{\pm} = \comm{X\sS}{\cdot}_\mp$. The super time-ordering operator $\overleftarrow{\mathcal{T}}$ chronologically orders only the super-operators.

The dissipation kernel $\kappa_r$ and noise kernel $\kappa_i$ are
\begin{eqnarray}
   \label{eq:kappa}
   \kappa_i(\tau) &=& \sum_j \frac{\nu_j^2}{2m_j\omega_j} \sin\omega_j\tau \\
   \kappa_r(\tau) &=& \sum_j \frac{\nu_j^2}{2m_j\omega_j} \coth\left(\frac{\omega_j\beta}{2}\right)\cos\omega_j\tau  
\end{eqnarray}
with $m_j$ and $\omega_j$ are mass and frequency of the $j$-th harmonic oscillator. The coupling strength  $\nu_j$ is defined in Eq. \eqref{eq:YB}.  When the Drude-Lorentz spectrum \eqref{eq:Drude} is used, the kernels decay multi-exponentially.\cite{Xu2009}  At a relatively high temperature, which we assumed for the environment, the kernels can be expressed as
\begin{equation}
   \kappa_r(\tau) - i \kappa_i (\tau) = \lambda \left(c_1 e^{-\gamma_1 \tau}+c_2 e^{-\gamma_2 \tau} + 2 c_0 \delta(\tau) \right).
\end{equation}
The constants $c_j$ and $\gamma_j$ can be obtained by various fitting methods. We used the values given in \cite{Tian2010}.

Using the influence functional, the system density operator at $t$ can be written as $\rho\sS(t) = e^{i \Phi(t,t_0)} \rho\sS(t_0)$.  However, applying the super-operator on the initial density is still a major difficulty in the influence functional approach.   HEOM avoids the difficulty of the exponential super-operator.  By differentiating $\rho\sS(t)$ with respect to time $t$, we find 
\begin{equation}
   \dv{t}\rho\sS(t)  =- i  \mathcal{S}^{-}(t) \left [ \left( \zeta_{1,0}(t) + \zeta_{0,1}(t)\right) - i c_0  \mathcal{S}^{-}(t) \rho\sS(t) \right ]
\end{equation}
where we introduced two auxiliary operators
\begin{equation}
   \begin{split}\label{eq:zeta10}
   \zeta_{1,0}(t) &= -i \overleftarrow{\mathcal{T}}\left ( \int_{t_0}^t \dd{s} e^{-\gamma_1 (t-s)} \mathcal{G}_1(s) \right) \\
   & \quad \times \prod_{j=1,2} \exp\left [ -\lambda \int_{t_0}^t \dd{s_1} \int_{t_0}^{s_1} \dd{s_2}  \right . \\ 
   & \left . \qquad \times \mathcal{S}^{-}(s_1) e^{-\gamma_j (s_1-s_2)} \mathcal{G}_j(s_2) \right ] \rho\sS(0)
   \end{split}
\end{equation}
\begin{equation}
   \begin{split}\label{eq:zeta01}
   \zeta_{0,1}(t) &= -i  \overleftarrow{\mathcal{T}}\left ( \int_{t_0}^t \dd{s} e^{-\gamma_2 (t-s)} \mathcal{G}_2(s) \right)  \\
   & \quad \times \prod_{j=1,2} \exp\left [ -\lambda \int_{t_0}^t \dd{s_1} \int_{t_0}^{s_1} \dd{s_2}  \right . \\
   & \left . \qquad \times \mathcal{S}^{-}(s_1) e^{-\gamma_j (s_1-s_2)} \mathcal{G}_j(s_2) \right ] \rho\sS(0)
   \end{split}
\end{equation}
and another super-operator
\begin{equation}
   \mathcal{G}_j(t) = \Re\{c_j\} \mathcal{S}^{-}(t) + i \Im\{c_j\} \mathcal{S}^{+}(t) 
\end{equation}

In order to find $\zeta_{0,1}$ and  $\zeta_{1,0}$, we differentiate Eqs. (\ref{eq:zeta10}) and (\ref{eq:zeta01}) which leads to a new set of auxiliary operators. By repeating the differentiation, we obtain a hierarchy of auxiliary operators
\begin{eqnarray}\label{eq:zeta_n1n2}
  \zeta_{n_1,n_2}(t) &=& \overleftarrow{\mathcal{T}} \left ( -i \int_{t_0}^t \dd{s} e^{-\gamma_1 (t-s)} \mathcal{G}_1(s) \right )^{n_1} \nonumber \\
  && \times \left ( -i \int_{t_0}^t \dd{s}  e^{-\gamma_2 (t-s)} \mathcal{G}_2(s) \right )^{n_2}  \nonumber\\
  && \times \exp\left [ -\lambda c_0 \int_{t_0}^t \dd{s} \mathcal{S}^{-}(s) \mathcal{S}^{-}(s) \right ] \nonumber\\
  && \times \prod_j \exp\left [ -\lambda \int_{t_0}^t \dd{s_1} \int_{t_0}^{s_1} \dd{s_2}  \mathcal{S}^{-}(s_1) \right.\nonumber \\
  && \left.  \times e^{-\gamma_j (s_1-s_2)} \mathcal{G}_j(s_2) \right ] \rho\sS(0)
\end{eqnarray}
which are determined by a  hierarchy of differential equations (now in the Schro\"odinger picture):
\begin{equation}
   \begin{split}\label{eq:HEOM-schroedinger}
  & \dv{t}\zeta_{n_1,n_2}(t) = -i \comm{H\sS}{\zeta_{n_1,n_2}(t)}_{-} \\
  & \quad - (\gamma_1 n_1 + \gamma_2 n_2) \zeta_{n_1,n_2}(t) - \lambda\, c_0\, \mathcal{S}^{-} \mathcal{S}^{-}\, \zeta_{n_1,n_2}(t) \\
  & \quad -i n_1 \mathcal{G}_1\, \zeta_{n_1-1,n_2}(t) -i n_2  \mathcal{G}_2\, \zeta_{n_1,n_2-1}(t) \\
  & \quad -i \lambda \,  \mathcal{S}^{-} \left\{\zeta_{n_1+1,n_2}(t) + \zeta_{n_1,n_2+1}(t) \right\}.
   \end{split}
\end{equation} 

Since the super-operators $S^{\pm}$ directly act on the Liouville space, we can evaluate the right hand side of Eq. (\ref{eq:HEOM-schroedinger}) without any difficulty.  Now the problem of exponential super-operator is replaced with an infinite set of ODEs.  The system density is at the top of the hierarchy $\rho\sS(t)=\zeta_{0,0}(t)$ and the moment operator $\eta_1(t)$ used in Eq. \eqref{eq:eom} is obtained from the auxiliary operators as
\begin{equation}
   \eta_1(t) =  \left[ \zeta_{1,0}(t)+\zeta_{0,1}(t) - i c_0 \mathcal{S}^{-}\zeta_{0,0}(t) \right ].
\end{equation}
Going down the hierarchy the contribution of deeper levels to the top three auxiliary operators becomes negligible and the hierarchy can be terminated at a certain depth without losing the accuracy of $\rho\sS$ and $\eta_1$.  The cutoff depth depends on the coupling strength.   For a system strongly coupled with environments, we must include many auxiliary operators of higher depth.  In the present simulation, the depth $d=50 \sim 70$ is found to be enough.

\bibliographystyle{apsrev4-2}
\bibliography{references}

\end{document}